\documentclass[
twocolumn,
]{ceurart}

\usepackage{listings}
\usepackage{xcolor}
\usepackage{makecell}
\usepackage{caption} 
\usepackage{tablefootnote}
\usepackage{algorithm}
\usepackage{enumitem}
\usepackage{algpseudocode}
\newcommand\numberthis{\addtocounter{equation}{1}\tag{\theequation}}
\begin{document}

%\copyrightyear{2023}

% \copyrightclause{Copyright for this paper by its authors.
%   Use permitted under Creative Commons License Attribution 4.0
%   International (CC BY 4.0).}

%\conference{Woodstock'22: Symposium on the irreproducible science,
%  June 07--11, 2022, Woodstock, NY}
% \conference{IJCAI 2023 Workshop on Deepfake Audio Detection and Analysis (DADA 2023), August 19, 2023, Macao, S.A.R}

% INTEGRATION OF FRAME-LEVEL BOUNDARY DETECTION AND DEEPFAKE DETECTION FOR THE LOCALIZATION OF AUDIO FORGERY MANIPULATED REGIONS
\title{THE DKU-DUKEECE SYSTEM FOR THE MANIPULATION REGION LOCATION TASK OF ADD 2023}

\author[1]{Zexin Cai}[%
email=zexin.cai@duke.edu,
]
\author[1]{Weiqing Wang}[%
email=weiqing.wang@duke.edu,
]
\address[1]{Department of Electrical and Computer Engineering, Duke University, Durham, NC 27708, USA}

\author[2]{Yikang Wang}[%
email=yikang.wang772@dukekunshan.edu.cn,
]

\author[2]{Ming Li}[%
email=ming.li369@duke.edu,
]
\cormark[1]

\address[2]{Data Science Research Center, Duke Kunshan University, Kunshan 215316, PR China}

\cortext[1]{Corresponding author.}
% \fntext[1]{These authors contributed equally.}


  % This paper introduces our system designed for locating manipulated regions in the second Audio Deepfake Detection Challenge (ADD 2023). Our approach involves the utilization of multiple detection systems to identify these regions and determine their authenticity by distinguishing between bonafide audio clips and synthesized ones. Specifically, we train and integrate two frame-level systems: one for boundary detection and the other for deepfake verification. Additionally, we employ a third VAE model trained exclusively on bonafide data to determine the authenticity of a given audio clip. By combining these three systems, our top-performing solution for the ADD challenge achieves an impressive 82.23\% sentence accuracy and an $F_1$ score of 67.13\%.
  
\begin{abstract}
This paper introduces our system designed for Track 2, which focuses on locating manipulated regions, in the second Audio Deepfake Detection Challenge (\textit{ADD 2023}). Our approach involves the utilization of multiple detection systems to identify splicing regions and determine their authenticity. Specifically, we train and integrate two frame-level systems: one for boundary detection and the other for deepfake detection. Additionally, we employ a third VAE model trained exclusively on genuine data to determine the authenticity of a given audio clip. Through the fusion of these three systems,  our top-performing solution for the ADD challenge achieves an impressive 82.23\% sentence accuracy and an $F_1$ score of 60.66\%. This results in a final ADD score of 0.6713, securing the first rank in Track 2 of \textit{ADD 2023}.
\end{abstract}

\begin{keywords}
  Audio splicing forgery \sep
  waveform boundary detection \sep
  audio deepfake detection challenge
\end{keywords}

\maketitle

\section{Introduction}
The increasing application of deep generation in diverse fields such as text, image, and audio has recently become accessible to the public with its human-like generative outputs~\cite{openai2023gpt4, rombach2022high, kim2021conditional}. However, such applications inherently carry a potential threat to society if misused. In addition, while the generation of highly realistic content is advancing rapidly in deep learning, the development of corresponding countermeasures for detecting fraudulent content is comparatively slow and significantly lagging behind. In the field of speech signal processing, as synthesized speech approaches the naturalness of human speech, there is a growing demand for detection systems capable of identifying synthesized speech. This demand has captured the attention of researchers and has prompted the organization of speech anti-spoofing challenges aimed at encouraging the development of effective countermeasures~\cite{wu2017asvspoof, 9746939}. 

Generally, speech synthesis comprises two techniques, text-to-speech and voice conversion. Different deep architectures, including Tacotron-based models~\cite{shen2018natural, Cai2020}, Fastspeech-based models~\cite{ren2022fastspeech, cai2023cross}, and VAE-based models~\cite{kim2021conditional}, have been proposed to achieve high-fidelity speech synthesis for those two synthesis methods. In this context, spoofing attacks can be launched through any of these approaches.  While the literature has reported impressive performance in anti-spoofing detection, most of these deep learning models are trained and evaluated on specific systems and datasets~\cite{kamble2020advances}. As a result, the discriminatory ability of these deep models significantly diminishes when confronted with unseen scenarios and mismatched domains~\cite{wang21_asvspoof}. For instance, detecting attacks generated by synthesis approaches that are not encountered in the training data. Therefore, the development of an anti-spoofing detection system that performs well on unseen scenarios is crucial.

With regards to the expanding usage of numerous synthesis systems and the challenges encountered in anti-spoofing and deepfake speech detection, the second Audio Deepfake Detection Challenge (\textit{ADD 2023}) \footnote{http://addchallenge.cn/add2023} has been organized to motivate researchers to explore new and innovative technologies that can accelerate and facilitate research in detecting and analyzing deepfake speech utterances \cite{add2023}. Track 2 of \textit{ADD 2023} focuses on the identification of partially spoofed speech, commonly known as audio splicing forgery. These speech attacks involve manipulated audio clips comprising real or synthesized audio segments. Unlike Track 2 in \textit{ADD 2022}~\cite{9746939}, which solely focused on determining the authenticity of given speech clips, \textit{ADD 2023} sets the objective of locating the fake regions as well, making the task considerably more challenging. Furthermore, the testing dataset for \textit{ADD 2023} contains out-of-domain data that has genuine speech from mismatch domains and synthetic audio clips generated using models distinct from those employed in the training set.

In this paper, we present our submitted system for Track 2 in the \textit{ADD 2023} challenge. Specifically, we train a waveform boundary detection system using the provided training data~\cite{cai2022waveform}. While the boundary detection system effectively identifies the boundaries of spliced audio and locates each segment, it does not determine the authenticity of individual audio segments. To address this limitation, we incorporate a frame-level anti-spoofing system to classify each located region as genuine or fake. We change the output labelling strategy of the boundary detection system and transform the system to a frame-level anti-spoofing detection system, which aims to verify the authenticity of each input audio frame. Both systems are built upon large-scale self-supervised pre-training models~\cite{baevski2020wav2vec, chen2022wavlm}, ensuring robust performance and reliable results. Furthermore, we introduce a VAE-based model trained exclusively on genuine audio samples from the training set. This model serves as a supplementary verification system, capable of identifying whether a given audio clip is an outlier relative to genuine audio samples. By combining these systems, we achieve a final score of 0.6713 in the ADD challenge. Notably, our system achieves an impressive sentence accuracy of 82.23\% and an $F_1$ score of 60.66\%.

% In Section~\ref{sec:relwork}, we provide an overview of related works from the literature.
The rest of the paper is structured as follows: Section~\ref{sec:method} outlines our proposed approach for detecting and locating fake regions. We elaborate on our experimental processes and analyze the result in Section~\ref{sec:exp}. Finally, in Section~\ref{sec:conclu}, we conclude our work. 

\begin{figure}[h!]
  \centering
  \includegraphics[width=0.35\textwidth]{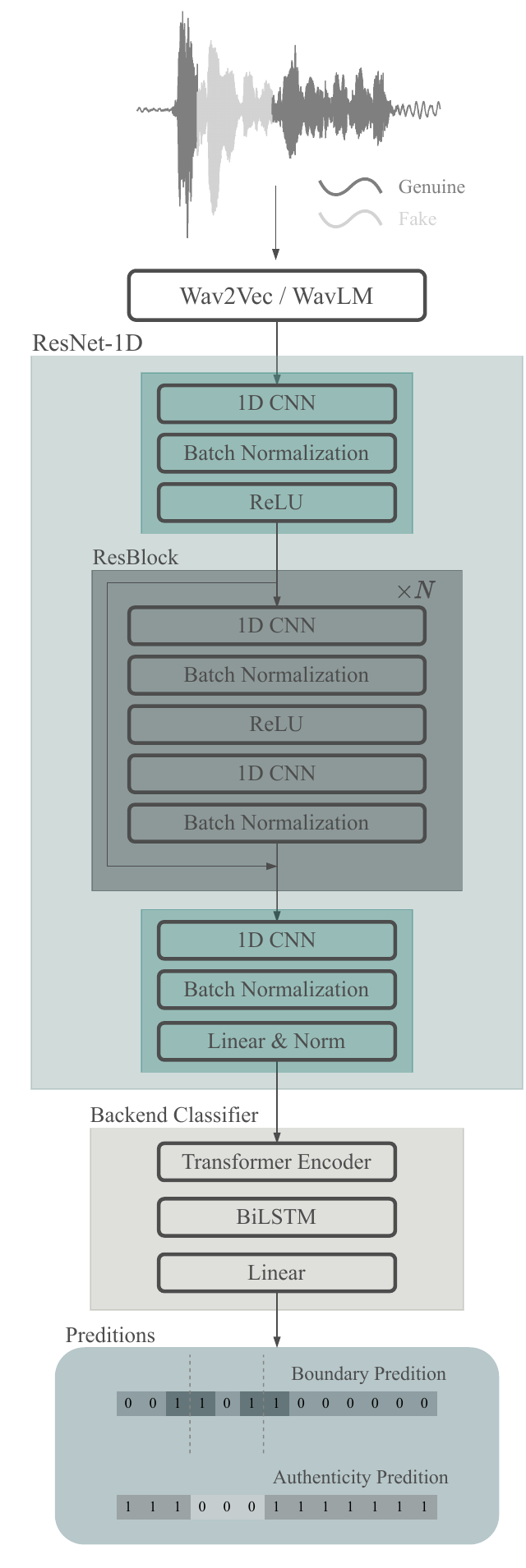}
  \caption{The overall architecture of frame-level detection models.}
  \label{fig:archi}
\end{figure}

% \section{Related works}
% \label{sec:relwork}

\section{Methods}
\label{sec:method}

Our system for the manipulated region location task incorporates three models. Two of these models share a similar network architecture with our boundary detection system for ADD 2022~\cite{cai2022waveform}, differing mainly in their output labelling strategy. The third model is a VAE-based classification model used as a supplementary component to assess the authenticity of a given audio clip.

\subsection{Frame-level detection models}
\label{sec:FLmodel}
The frame-level detection models used in our approach, one for boundary detection and the other for anti-spoofing detection, utilize the deep framework depicted in Figure~\ref{fig:archi}.To extract frame-level acoustic representations from raw audio signals, we employ large-scale self-supervised pre-training models such as Wav2Vec~\cite{wav2vec2} and WavLM~\cite{chen2022wavlm}. Following that, a 1-dimensional residual network module (ResNet-1D) is utilized to further extract frame-level features specific to our task. A backend classifier is then applied to predict the classification result for each frame. Specifically, the ResNet-1D module consists of two 1-dimensional convolutional neural network (1D-CNN) layers surrounding a series of residual blocks. Each residual block contains two 1D-CNN layers and incorporates a residual connection from input to output. The backend classifier incorporates transformer encoders and a Bidirectional LSTM (BLSTM) to capture long-range global contexts. It also includes a fully connected layer that maps the high-dimensional output vectors to binary outputs.

The boundary detection model aims to identify discontinuities in a sequence of frames. Consequently, the labelling strategy assigns a value of 1 to frames surrounding waveform concatenation boundaries, while frames within genuine and fake segments are labelled as 0. Conversely, the anti-spoofing detection model assigns a label of 0 to fake audio frames and 1 to genuine frames in the target output. Examples of the labelling process are illustrated in Figure~\ref{fig:archi}.

\subsection{VAE model}
The Variational Autoencoder (VAE) is a probabilistic graphical model that effectively reduces dimensionality in a statistically grounded manner. Unlike other techniques like autoencoders and principal component analysis (PCA), the VAE offers reconstructed probability as a measure of deviation rather than relying on reconstruction error as an anomaly score~\cite{an2015variational}. The reconstruction probability is commonly employed as the final score for deviation-based outlier detection~\cite{kingma2013auto}. Wang et al. demonstrated the applicability of this model in spoofing detection using real speech differential features as input~\cite{wang21_asvspoof}.

\section{Experiment}
\label{sec:exp}

\subsection{Data preparation}

Table~\ref{tab:dataset} provides an overview of the datasets provided by \textit{ADD 2023}, including an additional dataset called ADD-Eval that we generat for evaluating our models. The statistics in the table reveal the composition of each dataset. The ADD-Train dataset consists of 26,554 genuine utterances, 1,185 fake utterances, and 25,354 manipulated fake utterances. The ADD-Dev dataset contains 8,914 genuine utterances, 430 fake utterances, and 17,824 partially fake utterances. The ADD-test dataset consists of 50,000 unlabelled utterances without labelling information.

\begin{table}[h]
  % \captionsetup{justification=centering}
  \caption{The statistics of datasets (\#Utterances)}
  \label{tab:dataset}
  \centering
  \begin{tabular}[c]{ccccc}
    \toprule
    \textbf{Name}  & Genuine & Fake & PartialFake & All \\
    \midrule
    ADD-Train & 26,554 & 1,185 & 25,354 & 53,093\\
    ADD-Dev & 8,914 & 430 & 8,480 & 17,824 \\
    ADD-Test & - & - & - & 50,000 \\
    ADD-Eval & 4,000 & 4,000 & 6,000 & 14,000 \\
    \bottomrule
    
  \end{tabular}
\end{table}

The ADD-Eval is constructed as an out-of-domain dataset from the ADD-Dev using segments re-synthesis by the World vocoder~\cite{Morise2016WORLD}. The re-synthesis process involves analyzing audio clips with World to obtain the corresponding acoustic parameters, and then synthesizing new clips or segments using these parameters. The ADD-Eval dataset includes 4,000 randomly selected genuine utterances from ADD-Dev. Among the 4,000 fake utterances, 430 are taken directly from ADD-Dev, and an additional 430 are re-synthesized from these fake utterances. The remaining 3,140 utterances are re-synthesized by World using genuine utterances randomly selected from ADD-dev. For the partially fake utterances in ADD-Eval, 2,000 of them are sourced from ADD-Dev, while another 2,000 are re-synthesized by World, focusing only on re-synthesizing the fake clips within the utterances. The remaining 2,000 utterances are obtained by randomly slicing and concatenating genuine utterances from ADD-Dev, followed by re-synthesizing randomly selected segments.

\subsection{Training and inference}
The data loading process contains several scenarios. When sampling training data for the anti-spoofing detection model, we employ three strategies to cut the data to a fixed length $l$, with the following probability distribution: \{Strategy 1: 0.35, Strategy 2: 0.3, Strategy 3: 0.35\}. $l$ is set to 1.28 according to the result in~\cite{cai2022waveform}.

\begin{enumerate}
  \item Randomly choose genuine and fake utterances from the training set, with a probability of 0.3 for selecting genuine utterances and a probability of 0.7 for selecting fake utterances. Then cut to fixed length $l$ as input for training.
  \item Randomly segment and concatenate two genuine utterances from the training set. The resulting concatenated utterance has a length equal to the fixed length $l$.
  \item Randomly segment clips from the partially fake utterances in the training set.
\end{enumerate}

For the boundary detection models, the genuine utterances and fake utterances are considered as identical since there are no boundaries within these utterances. Consequently, the sampling process in Strategy 1 for boundary detection does not involve selecting between genuine and fake utterances. The distribution for data sampling in boundary detection models is as follows: \{Strategy 1: 0.2, Strategy 2: 0.4, Strategy 3: 0.4\}. Regarding label generation, we assign labels of 1 to the four closest frames surrounding each boundary.

We apply online data augmentation using the MUSAN~\cite{MUSAN} and RIRs~\cite{RIRs} corpus for all models. The models are trained using binary cross-entropy loss and the Adam optimizer for 200 epochs. The mini-batch size is set to 64, except for models based on wavLM features, where it is set to 16. The initial learning rate is $10^{-4}$, and we utilize the Noam scheduler~\cite{vaswani2017attention} with 1600 warm-up steps. 

For additional information regarding the configuration and hyperparameter settings, please refer to Table~\ref{table:architecture}. Since we employ a fixed length of 1.28 seconds during training, the input size $L$ in the table corresponds to 20,480. The number of frames $T$ is set to 64, considering that both Wav2Vec and WavLM operate at a frame rate of 20ms. For Wav2Vec-based models, we adopt Wav2Vec 2.0 (wav2vec2-base-960h\footnote{https://huggingface.co/facebook/wav2vec2-base-960h}) model which is pretrained on the Librispeech dataset. WavLM-based model use the WavLM Large\footnote{https://github.com/microsoft/unilm/tree/master/wavlm} as the acoustic feature extractor. 

For anti-spoofing models, we evaluate the $F_1$ score on the ADD-Dev set, while for boundary detection models, we evaluate the equal error rate (EER). For Wav2vec-based models, we take the average of the top five models with the lowest EER or highest $F_1$ score for inference and evaluation. However, we noticed a decline in performance when averaging models for wavLM-based models. Consequently, for wavLM-based models, we only consider the one with the best performance for evaluation purposes.

\begin{table*}[th]
    \caption{The network configuration of frame-level detection models, where $\mathbf{C}$(kernal size, padding, stride) denotes the convolutional layer, $\left[\cdot \right]$ denotes the residual block, $\mathbf{E}$(number of layers, number of heads, FFN size) denotes the Transformer Encoder, BiLSTM(number of layers, hidden units) denotes BiLSTM layer, Linear(input size, output size) denotes the fully-connected layer; $L$ relates to the duration of the input audio signal and $T$ is the number of label frames.}\

    \centering
    \begin{tabular}[c]{@{\ \ }l@{\ \ }c@{\ \ }c@{\ \ }c@{\ \ }c@{\ \ }}
        \toprule
        \textbf{Layer} & \textbf{Output Size} & \textbf{Structure} & \textbf{\#Parameters}  & \textbf{Others}\\
        \midrule
        Input audio & $L \times 1$ & - & -  & 1.28 seconds \\
        \midrule
        Wav2Vec & $T \times 768$ & - & 94.37M & wav2vec2-base-960h \\
        For WavLM & $T \times 1024$ & - & 316.62M & WavLM Large  \\
        \midrule
        1D-CNN & $T \times 512$ & $\mathbf{C}(5, 2, 1) \mathrm{w/o\ bias}$ & \makecell{1.97M (Wav2Vec) \\ 2.62M (WavLM)} & - \\
        \midrule
        ResBlock(s)  & $T \times 512$ & $\begin{bmatrix}
            \mathbf{C}(1, 0, 1)\ \mathrm{w/o\ bias} \\
            \mathbf{C}(1, 0, 1)\ \mathrm{w/o\ bias}
        \end{bmatrix}$ & 6.32M & 12 blocks  \\
        \midrule
        1D-CNN & $T \times 128$ & $\mathbf{C}(1, 0, 1)$ & 0.066M & -  \\
        \midrule
        Linear \& Norm & $T \times 128$ & Linear(128, 128) & 0.017M & -  \\
        \midrule
        Transformer Encoder & $T \times 128$ & $\mathbf{E}(2, 4, 1024)$ & 0.66M & dropout=0.2  \\
        \midrule
        BiLSTM & $T \times 128$ & BiLSTM(1, 128) & 0.26M & dropout=0   \\
        \midrule
        Linear (Boundary) & $T \times 1$ & Linear(256, 1) & - & Binary Cross Entropy Loss   \\
        For (Anti-spoofing) & $T \times 2$ & Linear(256, 2) & - & Cross Entropy Loss  \\
        \bottomrule
    \end{tabular}
    \label{table:architecture}
\end{table*}

For the VAE model, we utilize the open-source module pyod\footnote{https://github.com/yzhao062/pyod} to implement it. The encoder and decoder of the VAE model are constructed with three hidden layers each, with dimensions of 128, 64, and 32, respectively. We adopt feature extracted by WavLM and applied PCA for dimensionality reduction to obtain the final frame-level feature for the VAE model. During the dimensionality reduction process, we preserve 98\% of the energy to determine the ultimate dimension. In the training phase, we exclusively employ genuine samples from the training set to estimate the latent space distribution.

During the inference stage, we begin by dividing the audio signal into overlapping audio clips. Each clip has the same length as the training samples, which is 1.28 seconds, with a step size of 0.64 seconds. After obtaining the probabilities for each frame within the audio clips, we merge the results of all the clips by averaging the overlapping regions. For the VAE model, we compute the average frame-level output probabilities of the reconstruction function and utilize this value as the score for each sample.

\subsection{Results}
\label{sec:res}

Instead of relying on the Equal Error Rate (EER) for evaluation, Track 2 incorporates multiple metrics to calculate the final score of a submission. These metrics include sentence accuracy ($A$) and segment $F_1$ score, as formulated in Equation~\ref{att_eq1} and Equation~\ref{att_eq2}, respectively. In these equations, $TP$ represents true positive, $FP$ represents false positive, and $FN$ represents false negative. We also have a $F_1^*$ score solely used for the ADD-Test set. This score is computed with the same equation as $F_1$ while it takes ``fake" as positive and ``genuine" as negative. The final ADD score is computed as a weighted sum of the sentence-level accuracy $A$ and the segment $F_1^*$ score, as illustrated in Equation ~\ref{att_eq3}. 

\begin{gather*}
    A = \frac{\textit{\# utterances with correct classification}}{\textit{\# total utterances}} \numberthis \label{att_eq1} \\
    F_{1} = \frac{2 * TP}{2*TP + FN + FP} \numberthis \label{att_eq2} \\
    Score = 0.3 \times A + 0.7 \times F^*_1 \numberthis \label{att_eq3} \\
\end{gather*}

Table~\ref{tab:len_exp} presents the performance of our models on the ADD-Dev and ADD-Eval datasets. In addition to the regular WavLM model, we also have a variant that follows a different training strategy. This strategy involves freezing the WavLM model for the first 50 epochs and then unfreezing it for the remaining 150 epochs during training. However, we have observed that this strategy does not yield favorable results for the Wav2Vec-based models. When the parameters are frozen, the performance of Wav2Vec models remains unsatisfactory during training. Since sentence accuracy is correlated with Equal Error Rate (EER), we utilize EER as the evaluation metric when ground truth labels are available. The table demonstrates that the WavLM-based model achieves the best performance in the boundary detection task, achieving an EER of 0.09\% on the ADD-Dev set and 0.67\% on the ADD-Eval set. Similar trends are observed in the anti-spoofing system. However, when evaluating the out-of-domain dataset, ADD-Eval, the Wav2Vec-based model outperforms other models with an $F_1$ score of 0.4628.

\begin{table}[h]
  \captionsetup{justification=centering}
  \caption{Performance on ADD-Dev and ADD-Eval, where BDR denotes boundary detection, and SPF denotes anti-spoofing detection.}
  \label{tab:len_exp}
  \centering
  \begin{tabular}[c]{ccc}
    \toprule
    \textbf{BDR System} & \textbf{Dev (EER)} & \textbf{Eval (EER)} \\
    \midrule
    Wav2Vec  & 0.39\% & 1.98\%  \\
    WavLM & \textbf{0.09\%} & \textbf{0.67}\%  \\
    WavLM$^\ast$ & 0.14\%  & 1.325\%   \\
    \toprule
    \textbf{SPF System} & \textbf{Dev (EER / $F_1$)} & \textbf{Eval (EER / $F_1$)} \\
    \midrule
    Wav2Vec & 0.07\% / 0.9991 & \textbf{10.45\% / 0.4628} \\
    WavLM & \textbf{0.0\% / 0.9991} & 20.24\% / 0.4376   \\
    WavLM$^\ast$ & 0.01\% / 0.9991 & 20.88\% / 0.438 \\
    \bottomrule
    \multicolumn{3}{l}{$^\ast$The model is frozen for the first 50 epochs and then  }\\
    \multicolumn{3}{l}{unfrozen for the remaining 150 epochs. }\\
  \end{tabular}
\end{table}

The performance of individual models on the ADD-Test dataset is presented in Table~\ref{tab:exp_test}. The boundary detection systems are designed to identify concatenation boundaries but lack the ability to determine whether a segment is genuine or fake. Therefore, we only report the sentence accuracy for the boundary detection systems. Among these systems, the WavLM-based model, which had its parameters frozen for 50 epochs, achieved the highest performance with a sentence accuracy of 0.7747. 
In this scenario, all utterances that do not have detected boundaries are considered genuine. As a result, the fake utterances in the test set are treated as genuine if the model's output indicates the absence of boundaries within the utterance.

\begin{table}[h]
  \captionsetup{justification=centering}
  \caption{Performance on the ADD-Test, where BDR denotes boundary detection, and SPF denotes anti-spoofing detection.}
  \label{tab:exp_test}
  \centering
  \begin{tabular}[c]{c|c|c}
    \toprule
    \multirow{2}{*}{\textbf{Feature} } & \textbf{BDR System} & \textbf{SPF System} \\
    & Sentence Acc (A) & $F^*_1$ score \\
    \midrule
    Wav2Vec  & 0.7131 & 0.3426  \\
    WavLM & 0.7631 & \textbf{0.3446}  \\
    WavLM$^\ast$ & \textbf{0.7747}  & 0.3437   \\
    \bottomrule
    \multicolumn{3}{l}{$^\ast$The model is frozen for the first 50 epochs and then  }\\
    \multicolumn{3}{l}{unfrozen for the remaining 150 epochs. }\\
  \end{tabular}
\end{table}

Regarding the anti-spoofing systems, the obtained $F^*_1$ scores are not satisfactory, resulting in low sentence accuracy. Therefore, we only report the $F^*_1$ scores in the table. In these SPF systems, the predictions are determined by a threshold of 0.95, classifying frames with scores below this threshold as fake. All three systems demonstrate similar performance, with the regular WavLM-based model achieving the highest $F^*_1$ score of 0.3446.

Regarding the advantages of individual systems, we combine the boundary detection system and the anti-spoofing system using a scoring strategy. The scoring algorithm is shown in Algorithm \ref{alg:cap}. Initially, for each utterance in the ADD-test dataset, we extract segmented audio clips using the boundaries identified by the boundary detection model. Then, we utilize the anti-spoofing detection system to determine the authenticity of each segment. If the number of segments is 1, indicating that no boundary is detected and the utterance is either bona fide or fake without any inserted audio clips, we determine the authenticity of the utterance by assessing the proportion of frames classified as fake, utilizing a spoofing detection model. The threshold ratio is set to 0.4. Consequently, if fewer than 40\% of frames are classified as fake, the utterance is categorized as bona fide; otherwise, it is labeled as fake. 

\begin{algorithm*}[th]
\caption{Scoring algorithm for ADD 2023 Track 2}\label{alg:cap}
\begin{algorithmic}
\Require segmented audio clips from the boundary detection model
\State FakeProportionRatio = 0.4 \Comment{The Proportion of fake frames predicted by spoofing detection model}
\If{$\#$Segments $=1$}
    \If{$\frac{\#\text{Fake frames}}{\#\text{Frames}} < \text{FakeProportionRatio}$}
        \State Classify as a bona fide segment
    \Else 
        \State Classify as a fake segment
    \EndIf
\ElsIf{$\#$Segments $=2$}
    \If{$\frac{\#\text{Fake frames}_{seg1}}{\#\text{Frames}_{seg1}} > \text{FakeProportionRatio} \text{ and } \frac{\#\text{Fake frames}_{seg1}}{\#\text{Frames}_{seg1}} > \frac{\#\text{Fake frames}_{seg2}}{\#\text{Frames}_{seg2}}$}
        \State Classify segment$_2$ as a bona fide segment and segment$_1$ as a fake segment
    \ElsIf{$\frac{\#\text{Fake frames}_{seg2}}{\#\text{Frames}_{seg2}} > \text{FakeProportionRatio} \text{ and } \frac{\#\text{Fake frames}_{seg2}}{\#\text{Frames}_{seg2}} > \frac{\#\text{Fake frames}_{seg1}}{\#\text{Frames}_{seg1}}$} 
        \State Classify segment$_1$ as a bona fide segment and segment$_2$ as a fake segment
    \Else
        \State Assign the segment with shorter length as fake and the other as bona fide \\ \Comment{Prior knowledge from the training set, most fake clips are with shorter length}
    \EndIf

\ElsIf{$\#$Segments $=3$}
    \State Classify the middle segment as fake and the other two as bona fide segments

\Else
    \For{each audio segment}
        \If{$\frac{\#\text{Fake frames}}{\#\text{Frames}} < \text{FakeProportionRatio}$}
            \State Classify as a bona fide segment
        \Else 
            \State Classify as a fake segment
        \EndIf
    \EndFor
\EndIf

\end{algorithmic}
\end{algorithm*}

In cases where the number of segments is 2, we apply the following criteria to classify a segment as fake: 1. The proportion of predicted fake frames within the segment exceeds that of the other segment; 2. This proportion is greater than the threshold ratio of 0.4. Alternatively, if these conditions are not met, we designate the segment with the shorter length as fake based on insights gained from the training set of ADD 2023. For utterances with 3 segments, we classify the middle segment as fake and the other two as bona fide segments, again relying on knowledge derived from the training set. Lastly, for utterances with more than 3 segments, we determine their predicted authenticity based on the proportion of predicted fake frames. By employing this scoring strategy, the fusion of the WavLM$^\ast$-based boundary detection model and the Wav2Vec-based anti-spoofing detection model achieves the best result, with an ADD score of 0.6538.

To further enhance the performance, we incorporate the VAE model into the scoring process. The scores of utterances with either 0 boundaries or more than 10 boundaries detected by the boundary detection systems are rescored using the VAE model. Any utterances among the top 45\% samples with the greatest deviation from the genuine training samples were classified as fake data. After applying the VAE scoring strategy, our final fusion system attained an ADD score of 0.6713, with an $F^*_1$ score of 0.6066 and a sentence accuracy of 0.8223.

\section{Conclusion}
\label{sec:conclu}

This paper introduces the DKU-DUKEECE system developed for Track 2 of the \textit{ADD 2023} challenge. The system comprises three distinct components, each trained to address a specific aspect of detecting manipulated regions in partially fake utterances. These components include a boundary detection model for identifying concatenation boundaries in spliced audio, a frame-level anti-spoofing detection model for determining the authenticity of individual audio frames, and a VAE model for identifying outliers with significant deviation from genuine data. By fusing these three models using well-defined scoring strategies, we achieve an impressive ADD score of 0.6713 on the ADD test dataset. This integration and fusion of different systems allow us to effectively detect and locate manipulated regions within the partially fake utterances.

% \begin{acknowledgments}
%   Thanks to
% \end{acknowledgments}

%%
%% Define the bibliography file to be used
\newpage
\bibliography{sample-ceur}

\begin{thebibliography}{22}
\expandafter\ifx\csname natexlab\endcsname\relax\def\natexlab#1{#1}\fi
\providecommand{\url}[1]{\texttt{#1}}
\providecommand{\href}[2]{#2}
\providecommand{\path}[1]{#1}
\providecommand{\DOIprefix}{doi:}
\providecommand{\ArXivprefix}{arXiv:}
\providecommand{\URLprefix}{URL: }
\providecommand{\Pubmedprefix}{pmid:}
\providecommand{\doi}[1]{\href{http://dx.doi.org/#1}{\path{#1}}}
\providecommand{\Pubmed}[1]{\href{pmid:#1}{\path{#1}}}
\providecommand{\bibinfo}[2]{#2}
\ifx\xfnm\relax \def\xfnm[#1]{\unskip,\space#1}\fi
%Type = Misc
\bibitem[{OpenAI(2023)}]{openai2023gpt4}
\bibinfo{author}{OpenAI}, \bibinfo{title}{{GPT-4 Technical Report}},
  \bibinfo{year}{2023}. \href{http://arxiv.org/abs/2303.08774}{{\tt
  arXiv:2303.08774}}.
%Type = Inproceedings
\bibitem[{Rombach et~al.(2022)Rombach, Blattmann, Lorenz, Esser, and
  Ommer}]{rombach2022high}
\bibinfo{author}{R.~Rombach}, \bibinfo{author}{A.~Blattmann},
  \bibinfo{author}{D.~Lorenz}, \bibinfo{author}{P.~Esser},
  \bibinfo{author}{B.~Ommer},
\newblock \bibinfo{title}{{High-resolution Image Synthesis with Latent
  Diffusion Models}},
\newblock in: \bibinfo{booktitle}{Proceedings of the IEEE/CVF Conference on
  Computer Vision and Pattern Recognition}, \bibinfo{year}{2022}, pp.
  \bibinfo{pages}{10684--10695}.
%Type = Inproceedings
\bibitem[{Kim et~al.(2021)Kim, Kong, and Son}]{kim2021conditional}
\bibinfo{author}{J.~Kim}, \bibinfo{author}{J.~Kong}, \bibinfo{author}{J.~Son},
\newblock \bibinfo{title}{{Conditional Cariational Autoencoder with Adversarial
  Learning for End-to-end Text-to-speech}},
\newblock in: \bibinfo{booktitle}{International Conference on Machine
  Learning}, \bibinfo{year}{2021}, pp. \bibinfo{pages}{5530--5540}.
%Type = Article
\bibitem[{Wu et~al.(2017)Wu, Yamagishi, Kinnunen, Hanil{\c{c}}i, Sahidullah,
  Sizov, Evans, Todisco, and Delgado}]{wu2017asvspoof}
\bibinfo{author}{Z.~Wu}, \bibinfo{author}{J.~Yamagishi},
  \bibinfo{author}{T.~Kinnunen}, \bibinfo{author}{C.~Hanil{\c{c}}i},
  \bibinfo{author}{M.~Sahidullah}, \bibinfo{author}{A.~Sizov},
  \bibinfo{author}{N.~Evans}, \bibinfo{author}{M.~Todisco},
  \bibinfo{author}{H.~Delgado},
\newblock \bibinfo{title}{{ASVspoof: The Automatic Speaker Verification
  Spoofing and Countermeasures Challenge}},
\newblock \bibinfo{journal}{IEEE Journal of Selected Topics in Signal
  Processing} \bibinfo{volume}{11} (\bibinfo{year}{2017})
  \bibinfo{pages}{588--604}.
%Type = Inproceedings
\bibitem[{Yi et~al.(2022)Yi, Fu, Tao, Nie, Ma, Wang, Wang, Tian, Bai, Fan,
  Liang, Wang, Zhang, Yan, Xu, Wen, and Li}]{9746939}
\bibinfo{author}{J.~Yi}, \bibinfo{author}{R.~Fu}, \bibinfo{author}{J.~Tao},
  \bibinfo{author}{S.~Nie}, \bibinfo{author}{H.~Ma}, \bibinfo{author}{C.~Wang},
  \bibinfo{author}{T.~Wang}, \bibinfo{author}{Z.~Tian},
  \bibinfo{author}{Y.~Bai}, \bibinfo{author}{C.~Fan},
  \bibinfo{author}{S.~Liang}, \bibinfo{author}{S.~Wang},
  \bibinfo{author}{S.~Zhang}, \bibinfo{author}{X.~Yan},
  \bibinfo{author}{L.~Xu}, \bibinfo{author}{Z.~Wen}, \bibinfo{author}{H.~Li},
\newblock \bibinfo{title}{{ADD 2022: The First Audio Deep Synthesis Detection
  Challenge}},
\newblock in: \bibinfo{booktitle}{IEEE International Conference on Acoustics,
  Speech and Signal Processing}, \bibinfo{year}{2022}, pp.
  \bibinfo{pages}{9216--9220}.
%Type = Inproceedings
\bibitem[{Shen et~al.(2018)Shen, Pang, Weiss, Schuster, Jaitly, Yang, Chen,
  Zhang, Wang, Skerrv-Ryan et~al.}]{shen2018natural}
\bibinfo{author}{J.~Shen}, \bibinfo{author}{R.~Pang}, \bibinfo{author}{R.~J.
  Weiss}, \bibinfo{author}{M.~Schuster}, \bibinfo{author}{N.~Jaitly},
  \bibinfo{author}{Z.~Yang}, \bibinfo{author}{Z.~Chen},
  \bibinfo{author}{Y.~Zhang}, \bibinfo{author}{Y.~Wang},
  \bibinfo{author}{R.~Skerrv-Ryan}, et~al.,
\newblock \bibinfo{title}{{Natural TTS Synthesis by Conditioning Wavenet on Mel
  Spectrogram Predictions}},
\newblock in: \bibinfo{booktitle}{IEEE International Conference on Acoustics,
  Speech and Signal Processing}, \bibinfo{year}{2018}, pp.
  \bibinfo{pages}{4779--4783}.
%Type = Inproceedings
\bibitem[{Cai et~al.(2020)Cai, Zhang, and Li}]{Cai2020}
\bibinfo{author}{Z.~Cai}, \bibinfo{author}{C.~Zhang}, \bibinfo{author}{M.~Li},
\newblock \bibinfo{title}{{From Speaker Verification to Multispeaker Speech
  Synthesis, Deep Transfer with Feedback Constraint}},
\newblock in: \bibinfo{booktitle}{Proc. Interspeech}, \bibinfo{year}{2020}, pp.
  \bibinfo{pages}{3974--3978}.
%Type = Misc
\bibitem[{Ren et~al.(2022)Ren, Hu, Tan, Qin, Zhao, Zhao, and
  Liu}]{ren2022fastspeech}
\bibinfo{author}{Y.~Ren}, \bibinfo{author}{C.~Hu}, \bibinfo{author}{X.~Tan},
  \bibinfo{author}{T.~Qin}, \bibinfo{author}{S.~Zhao},
  \bibinfo{author}{Z.~Zhao}, \bibinfo{author}{T.-Y. Liu},
  \bibinfo{title}{{FastSpeech 2: Fast and High-Quality End-to-End Text to
  Speech}}, \bibinfo{year}{2022}. \href{http://arxiv.org/abs/2006.04558}{{\tt
  arXiv:2006.04558}}.
%Type = Article
\bibitem[{Cai et~al.(2023)Cai, Yang, and Li}]{cai2023cross}
\bibinfo{author}{Z.~Cai}, \bibinfo{author}{Y.~Yang}, \bibinfo{author}{M.~Li},
\newblock \bibinfo{title}{{Cross-lingual Multi-speaker Speech Synthesis with
  Limited Bilingual Training Data}},
\newblock \bibinfo{journal}{Computer Speech \& Language} \bibinfo{volume}{77}
  (\bibinfo{year}{2023}) \bibinfo{pages}{101427}.
%Type = Article
\bibitem[{Kamble et~al.(????)Kamble, Sailor, Patil, and
  Li}]{kamble2020advances}
\bibinfo{author}{M.~R. Kamble}, \bibinfo{author}{H.~B. Sailor},
  \bibinfo{author}{H.~A. Patil}, \bibinfo{author}{H.~Li},
\newblock \bibinfo{title}{{Advances in Anti-spoofing: From the Perspective of
  ASVspoof Challenges}},
\newblock \bibinfo{journal}{APSIPA Transactions on Signal and Information
  Processing} \bibinfo{volume}{9} (????).
%Type = Inproceedings
\bibitem[{Wang et~al.(2021)Wang, Qin, Zhu, Wang, Zhang, and
  Li}]{wang21_asvspoof}
\bibinfo{author}{X.~Wang}, \bibinfo{author}{X.~Qin}, \bibinfo{author}{T.~Zhu},
  \bibinfo{author}{C.~Wang}, \bibinfo{author}{S.~Zhang},
  \bibinfo{author}{M.~Li},
\newblock \bibinfo{title}{{The DKU-CMRI System for the ASVspoof 2021 Challenge:
  Vocoder based Replay Channel Response Estimation}},
\newblock in: \bibinfo{booktitle}{Proc. 2021 Edition of the Automatic Speaker
  Verification and Spoofing Countermeasures Challenge}, \bibinfo{year}{2021},
  pp. \bibinfo{pages}{16--21}.
%Type = Inproceedings
\bibitem[{Yi et~al.(2023)Yi, Tao, Fu, Yan, Wang, Wang, Zhang, Zhang, Zhao, Ren,
  Xu, Zhou, Gu, Wen, Liang, Lian, Nie, and Li}]{add2023}
\bibinfo{author}{J.~Yi}, \bibinfo{author}{J.~Tao}, \bibinfo{author}{R.~Fu},
  \bibinfo{author}{X.~Yan}, \bibinfo{author}{C.~Wang},
  \bibinfo{author}{T.~Wang}, \bibinfo{author}{C.~Y. Zhang},
  \bibinfo{author}{X.~Zhang}, \bibinfo{author}{Y.~Zhao},
  \bibinfo{author}{Y.~Ren}, \bibinfo{author}{L.~Xu}, \bibinfo{author}{J.~Zhou},
  \bibinfo{author}{H.~Gu}, \bibinfo{author}{Z.~Wen},
  \bibinfo{author}{S.~Liang}, \bibinfo{author}{Z.~Lian},
  \bibinfo{author}{S.~Nie}, \bibinfo{author}{H.~Li},
\newblock \bibinfo{title}{{ADD 2023: The Second Audio Deepfake Detection
  Challenge}},
\newblock in: \bibinfo{booktitle}{The 5th IEEE International Workshop on Deep
  Analysis of Data-Driven Applications}, \bibinfo{year}{2023}.
%Type = Inproceedings
\bibitem[{Cai et~al.(2023)Cai, Wang, and Li}]{cai2022waveform}
\bibinfo{author}{Z.~Cai}, \bibinfo{author}{W.~Wang}, \bibinfo{author}{M.~Li},
\newblock \bibinfo{title}{Waveform boundary detection for partially spoofed
  audio},
\newblock in: \bibinfo{booktitle}{IEEE International Conference on Acoustics,
  Speech and Signal Processing}, \bibinfo{year}{2023}.
%Type = Article
\bibitem[{Baevski et~al.(2020)Baevski, Zhou, Mohamed, and
  Auli}]{baevski2020wav2vec}
\bibinfo{author}{A.~Baevski}, \bibinfo{author}{Y.~Zhou},
  \bibinfo{author}{A.~Mohamed}, \bibinfo{author}{M.~Auli},
\newblock \bibinfo{title}{{Wav2vec 2.0: A Framework for Self-supervised
  Learning of Speech Representations}},
\newblock \bibinfo{journal}{Advances in neural information processing systems}
  \bibinfo{volume}{33} (\bibinfo{year}{2020}) \bibinfo{pages}{12449--12460}.
%Type = Article
\bibitem[{Chen et~al.(2022)Chen, Wang, Chen, Wu, Liu, Chen, Li, Kanda,
  Yoshioka, Xiao et~al.}]{chen2022wavlm}
\bibinfo{author}{S.~Chen}, \bibinfo{author}{C.~Wang},
  \bibinfo{author}{Z.~Chen}, \bibinfo{author}{Y.~Wu}, \bibinfo{author}{S.~Liu},
  \bibinfo{author}{Z.~Chen}, \bibinfo{author}{J.~Li},
  \bibinfo{author}{N.~Kanda}, \bibinfo{author}{T.~Yoshioka},
  \bibinfo{author}{X.~Xiao}, et~al.,
\newblock \bibinfo{title}{{Wavlm: Large-scale Self-supervised Pre-training for
  Full Stack Speech Processing}},
\newblock \bibinfo{journal}{IEEE Journal of Selected Topics in Signal
  Processing} \bibinfo{volume}{16} (\bibinfo{year}{2022})
  \bibinfo{pages}{1505--1518}.
%Type = Article
\bibitem[{Baevski et~al.(2020)Baevski, Zhou, Mohamed, and Auli}]{wav2vec2}
\bibinfo{author}{A.~Baevski}, \bibinfo{author}{Y.~Zhou},
  \bibinfo{author}{A.~Mohamed}, \bibinfo{author}{M.~Auli},
\newblock \bibinfo{title}{{Wav2vec 2.0: A Framework for Self-supervised
  Learning of Speech Representations}},
\newblock \bibinfo{journal}{Advances in Neural Information Processing Systems}
  \bibinfo{volume}{33} (\bibinfo{year}{2020}) \bibinfo{pages}{12449--12460}.
%Type = Article
\bibitem[{An and Cho(2015)}]{an2015variational}
\bibinfo{author}{J.~An}, \bibinfo{author}{S.~Cho},
\newblock \bibinfo{title}{{Variational Autoencoder Based Anomaly Detection
  Using Reconstruction Probability}},
\newblock \bibinfo{journal}{Special lecture on IE} \bibinfo{volume}{2}
  (\bibinfo{year}{2015}) \bibinfo{pages}{1--18}.
%Type = Article
\bibitem[{Kingma and Welling(2013)}]{kingma2013auto}
\bibinfo{author}{D.~P. Kingma}, \bibinfo{author}{M.~Welling},
\newblock \bibinfo{title}{{Auto-encoding Variational Bayes}},
\newblock \bibinfo{journal}{arXiv preprint arXiv:1312.6114}
  (\bibinfo{year}{2013}).
%Type = Article
\bibitem[{Morise et~al.(2016)Morise, Yokomori, and Ozawa}]{Morise2016WORLD}
\bibinfo{author}{M.~Morise}, \bibinfo{author}{F.~Yokomori},
  \bibinfo{author}{K.~Ozawa},
\newblock \bibinfo{title}{{WORLD: A Vocoder-Based High-Quality Speech Synthesis
  System for Real-Time Applications}},
\newblock \bibinfo{journal}{IEICE Transactions on Information Systems}
  \bibinfo{volume}{E99} (\bibinfo{year}{2016}) \bibinfo{pages}{1877--1884}.
%Type = Article
\bibitem[{Snyder et~al.(2015)Snyder, Chen, and Povey}]{MUSAN}
\bibinfo{author}{D.~Snyder}, \bibinfo{author}{G.~Chen},
  \bibinfo{author}{D.~Povey},
\newblock \bibinfo{title}{{MUSAN}: {A} {Music}, {Speech}, and {Noise}
  {Corpus}},
\newblock \bibinfo{journal}{arXiv:1510.08484}  (\bibinfo{year}{2015}).
%Type = Inproceedings
\bibitem[{Ko et~al.(2017)Ko, Peddinti, Povey, Seltzer, and Khudanpur}]{RIRs}
\bibinfo{author}{T.~Ko}, \bibinfo{author}{V.~Peddinti},
  \bibinfo{author}{D.~Povey}, \bibinfo{author}{M.~L. Seltzer},
  \bibinfo{author}{S.~Khudanpur},
\newblock \bibinfo{title}{{A Study on Data Augmentation of Reverberant Speech
  for Robust Speech Recognition}},
\newblock in: \bibinfo{booktitle}{IEEE ICASSP}, \bibinfo{year}{2017}, pp.
  \bibinfo{pages}{5220--5224}.
%Type = Inproceedings
\bibitem[{Vaswani et~al.(2017)Vaswani, Shazeer, Parmar, Uszkoreit, Jones,
  Gomez, Kaiser, and Polosukhin}]{vaswani2017attention}
\bibinfo{author}{A.~Vaswani}, \bibinfo{author}{N.~Shazeer},
  \bibinfo{author}{N.~Parmar}, \bibinfo{author}{J.~Uszkoreit},
  \bibinfo{author}{L.~Jones}, \bibinfo{author}{A.~N. Gomez},
  \bibinfo{author}{{\L}.~Kaiser}, \bibinfo{author}{I.~Polosukhin},
\newblock \bibinfo{title}{{Attention Is All You Need}},
\newblock in: \bibinfo{booktitle}{{Advances in Neural Information Processing
  Systems}}, \bibinfo{year}{2017}, pp. \bibinfo{pages}{5998--6008}.

\end{thebibliography}


%\appendix


% \begin{itemize}
% \item \href{https://github.com/yamadharma/ceurart}{GitHub},
% % \item \href{https://www.overleaf.com/project/5e76702c4acae70001d3bc87}{Overleaf},
% \item
%   \href{https://www.overleaf.com/latex/templates/template-for-submissions-to-ceur-workshop-proceedings-ceur-ws-dot-org/pkfscdkgkhcq}{Overleaf
%     template}.
% \end{itemize}

\end{document}